\documentclass[12pt]{iopart}
\usepackage{graphicx} 
\begin{document}

\title	
	{Ternary configuration in the framework of inverse
	mean-field  method}

\author{
        V G  Kartavenko$^{a,b}$
,\
	A S\u{a}ndulescu$^{a,c}$\ and  W Greiner$^a$} 

\address{
  	$^a$\ Institut f\"ur Theoretische Physik der J. W. Goethe
  	Universit\"at, D-60054 Frankfurt am Main, Germany
  	}

\address{
  	$^b$\ Bogoliubov Laboratory of Theoretical Physics, 
  	Joint Institute for Nuclear Research,  
  	Dubna, 141980, Moscow District, Russia
  	} 

\address{
  	$^c$\ Romanian Academy, Calea Victoriei 125, Bucharest, 71102,
  	Romania
  	}

\begin{abstract}
A static scission configuration in cold ternary fission
has been considered in the framework of mean field approach.
The inverse scattering method is applied
to solve single-particle Schr\"odinger equation,
instead of constrained
selfconsistent Hartree--Fock equations.
It is shown,
that it  is possible to simulate one-dimensional three-center system 
via inverse scattering method
in the approximation of reflectless single-particle potentials.
\end{abstract}
\vskip -5mm


\maketitle

\section{Introduction} 
Ternary fission involving the emission of
$\alpha$-particle was first observed \cite{alvarez} more than 
fifty years ago. 
Emission of $\alpha$-particles in the spontaneous fission 
of $^{252}$Cf has also a long history of investigation 
experimentally \cite{fraenkel67} and  theoretically 
\cite{fraenkel67etc,sandulIJMPE99}.

A renewed interest in the spontaneous fission of $^{252}$Cf 
arose in connection with  modern experimental techniques, 
$\gamma-\gamma-\gamma$ and $x-\gamma-\gamma$ 
triple coincidence,
(Gammasphere with 110 Compton suppressed
Ge detectors), which allow 
the fine resolution of the mass, charge 
and angular momentum content of the fragments\cite{terakopian94}. 
Very recently direct experimental evidence was presented 
for the cold  (neutronless) ternary spontaneous fission of $^{252}$Cf 
in which the third particle is an $\alpha$-particle 
\cite{He4ternary}, or $^{10}$Be \cite{Be10ternary}.
This confirms that a large variety of nuclear
large-amplitude collective motions such as bimodal fission \cite{hulet86},
cold binary  fission 
\cite{terakopian94,Hamb93,Schwab94,Dard96,sandul96},  
heavy cluster radioactivity \cite{price89,spg80},
and inverse processes, such as subbarier fusion \cite{armbruster}, could
belong to the general phenomenon of cold nuclear fragmentation.

The main common characteristic of cold binary nuclear fragmentation is 
the emission of
the final fragments (nuclei) with very low 
or even zero excitation energy and consequently
with high total kinetic energy (TKE), which can be provided only with compact
shapes of fragments at the scission point \cite{sandul89,goenborsig90}.
The cold ternary fragmentations should also have compact shapes at scission
with deformations close to those of their ground states, in order to achieve
high TKE values tending to the ternary decay energy $Q_t$.
It is assumed, that a light nucleus (e.g. an alpha-particle, Be etc.) 
is formed at the surface of the initial
nucleus, and that the two heavier fragments are formed at the scission
configuration in their ground states.

These cluster like models \cite{sandul89,goenborsig90} 
were used successfully to reproduce general features of
the cold ternary fragmentation. However the scission configuration has been
built in fact by hands. Therefore it is actual to develop microscopical or
semi-microscopical approach to this scission-point concept of nuclear
fragmentation.

Recent experimental progress in the production of superheavy 
elements \cite{114118}
near the proposed superheavy "island of stability" \cite{hofmann98} 
has stimulated yet the critical reexamination of earlier extrapolations 
\cite{SuperNils,Mosel,Pat91a,Mol94a} of the nuclear shell structure.
Although this macroscopic-microscopic models quite successfully
describe the bulk properties of known nuclei, 
their parameterization needs preconceived knowledge about
the density distribution and the nuclear potentials which fades away
when going to the limits of stability.
Therefore it is actual to investigate the properties 
of superheavy elements with self-consistent models,
like the mean-field models based on the shell correction method,
self-consistent  Skyrme-Hartree-Fock (SHF) \cite{Vaut70}  and 
relativistic mean-field (RMF) models 
(see \cite{WG99june,Ring,Naz,WGijmpe95,Bender98} and refs. therein).


There are well developed methods to calculate, in the framework of
many-body self--consistent approach,
static properties
of a well isolated nucleus in its ground state.
There also exists a well developed two-center shell model \cite{2csh72}.
However, a three-center shell model has not been developed yet,
except for very early steps \cite{sg3c1977}. 
Three-center shapes are practically not
investigated, in comparison with the two-center ones.
There exists the  generalizations of mean--field models to the case of 
two-centers \cite{berger89}, but
a ternary configuration is out of consideration, 
because of uncertainties to select 
a peculiar set of constraints.

There also exist a number of calculations for nucleus-nucleus collisions 
in the frame of time-dependent mean--field methods, but an evolution of
the cold fragmentation has not been investigated yet.

Therefore, although the principal way to describe nuclear fragmentation in the
framework of many-body self-consistent approach exists, it is interesting to
develop other mean-field approaches to analyse these phenomena 
from different points of view.    

We suggest to use the methods of nonlinear dynamics.  
These methods
gave yet the possibility to derive
for nuclear physics unexpected collective modes,
which can not be obtained by traditional  methods of perturbation
theory near some equilibrium state (see 
\cite{kart93pn,KLSG96,KGMG96,KGG98,KSG98} for the recent refs.).
The most importamt reason is that 
the fragmentation and clusterization is a very general phenomenon. 
There are cluster objects 
in subnuclear and macro physics. 
Very different
theoretical methods were developed in these fields. However,
there are only few basic physical ideas, and most of the methods
deal with nonlinear partial differential equations.
One of the most important part of soliton theory is the inverse scattering
method \cite{gelfand51,marchenko55,faddev59}
and its applications to the integration of nonlinear partial
differential equations \cite{novikov70}.

In this Letter, we simulate three-center configuration
using inverse scattering method
to solve the single-particle Schroedinger equation, 
instead of direct solution of
constrained Hartree-Fock equations.
%
%
\section{The framework}
%
%
The inverse methods to integrate nonlinear evolution equations
are often more effective than a direct numerical integration.
Let us demonstrate this statement for the following simple case.
The type of systems under consideration are slabs of
nuclear matter \cite{bcn76}, which are finite in the $z$ coordinate and
infinite and homogeneous in two transverse directions.
The wave function for the slab geomethry is
\begin{equation}
\psi_{{{\mathbf k}_{\perp}} n}
({\mathbf x}) = {1 \over {\sqrt{\Omega}}} {\psi}_{n} (z)
\exp (i {\mathbf k}_{\perp}{\mathbf r}),\qquad
\epsilon_{{{\mathbf k}_{\perp}} n} = {\hbar^{2} k_{\perp} ^{2}
\over 2m} + e_{n}, 
\label{eq:slab}
\end{equation}
where ${\mathbf r} \equiv (x,y), \/ {{\mathbf k}_{\perp}}
\equiv (k_{x}, k_{y})$, and $\Omega$ is the transverse normalization area.\\
\begin{equation}
-{\hbar^{2} \over 2m} {d^{2} \over dz^{2}}
\psi_{n} (z) + U(z) \psi_{n} (z) = e_{n} \psi _{n} (z),
\label{eq:sp-problem}
\end{equation}
A direct method to solve the single-particle problem  (\ref{eq:sp-problem})
is to assign a functional of interaction ${\cal E}$
(usually an effective density dependent Skyrme force), 
to derive the ansatz for the one-body potential,
as the first variation of a functional of interaction in density
%
$U(z) = U[\rho(z)] = \delta{\cal E}/\delta\rho.$
%
Then to solve the Hartree-Fock problem under the set of constraints,
which define the specifics of the nuclear system.
In the simplest case of a ground state, one should conserve the total particle
number of nucleons ($A$), which is related to the "thickness" of a slab, via
$A \Longrightarrow {\cal A} = ( 6 A {\rho _{N} ^{2}} / \pi ) ^{1/3}$,
which gives the same radius for a three-dimensional system and
its one-dimensional analogue.
As a result, one obtains the energies of the single particle states
$e_{n}$, their wave functions $\psi _{n} (z)$, the density profile 
$\rho ({\mathbf x}) \Longrightarrow \rho (z)$
\begin{equation}
\rho (z) =
\sum _{n=1} ^{N_{0}} a_{n} \psi _{n} ^{2} (z),\qquad
{\cal A} = 
\sum _{n=1} ^{N_{0}} a_{n}, \qquad
a_{n} = {2m \over \pi \hbar^{2} } ( e_{F} - e_{n}),
\label{eq:density}
\end{equation}
and the corresponding single-particle potential.
$a_{n}$ are the occupation numbers, 
$ N_{0}$ is the number of  occupied bound orbitals. 
The Fermy-energy $e_{F}$ controls
the conservation of the total number of nucleons. 
%
%
The energy (per nucleon) of a system  is given by 
\begin{equation}
{E \over A} \Longrightarrow { \hbar^{2}
\over 2m{\cal A}} \Bigr(
 \sum _{n=1} ^{N_{0}} a _{n} \int _{-\infty} ^{\infty}
\bigr( {d \psi_{n} \over dz} \bigl) ^{2} dz
+ {\pi \over 2} \sum _{n=1} ^{N_{0}} a_{n} ^{2} \Bigl)
 + {1 \over {\cal A}} \int _{-\infty} ^{\infty}
{\cal E} [ \rho (z) ] dz.
\label{eq:energy}
\end{equation}
Finally, the set of formulas (\ref{eq:slab}--\ref{eq:energy}) completely 
defines the direct self-consistent problem.\\
Following the inverse scattering method, 
one reduces the main  differential Schr\"odinger equation
(\ref{eq:sp-problem}) to 
the integral Gel'fand-Levitan-Marchenko equation
\cite{gelfand51,marchenko55} 
\begin{equation}
K (x, y) + B (x + y) + \int _{x} ^{\infty} B (y + z) K (x,z) dz =0.
\label{eq:GLM}
\end{equation}
for a function $K(x,y)$.
The kernel $B$ is determined by the reflection coefficients
$R (k) \/ ( e_{k} = \hbar^{2} k^{2} / 2m )$,
and by the $N$ bound state eigenvalues
\[
B (z) = \sum _{n=1} ^{N} C_{n} ^{2} ( \kappa_{n} )  +
{1 \over \pi} \int _{-\infty} ^{\infty} R (k) \exp \/ (ikz)\/ dk, \qquad
e_{n} = - \hbar^{2} \kappa_{n} ^{2} / 2m .
\]
$N $ is the total number of the bound orbitals.
The coefficients $C _{n}$ are uniquely specified by the boundary conditions
and the symmetry of the problem under consideration.
The general solution,
%
$U (z) = - (\hbar^{2} / m) (\partial K (z, z)/ \partial z),$
%
should naturally contain both,
contributions due to the continuum of the spectrum and to its discrete
part. There seems to be no way to obtain 
the general solution $U(z)$ in a closed
form. Eqs.~(\ref{eq:sp-problem}),(\ref{eq:GLM}) 
have to be solved only numerically.
In Ref. \cite{kartmad87} we used the approximation of
reflectless ($R(k) = 0$),
symmetrical ($U(-z) = U(z)$) potentials.
This gave the possibility to derive
the following set of relations
\begin{eqnarray} \nonumber
U (z) = - {\hbar^{2} \over m} {\partial^{2} \over \partial z^{2}}
\ln ( \det \Vert M \Vert ) = - {2 \hbar^{2} \over m}
\sum _{n=1} ^{N} \kappa_{n} \psi _{n} ^{2} (z),\\ \nonumber
\psi _{n} (z) = \sum _{n=1} ^{N} ( M^{-1} )_{nl} \lambda _{l} (z),\qquad
\lambda _{n} (z) = C_{n} (\kappa_{n}) \exp \/ ( - \kappa_{n}z ),\\
M_{nl} (z) = \delta_{nl} + {{\lambda_{n} (z) \lambda_{l} (z)} \over
{\kappa_{n} + \kappa_{l}}},\qquad 
C_{n} (\kappa_{n}) = \Bigl( 2 \kappa_{n} 
\big{\vert} {\prod _{l {\not=} n} ^{N}}
{{\kappa_{n} + \kappa_{l}} \over {\kappa_{n} - \kappa_{l}}} 
\vert \Bigr) ^{1/2}.
\label{eq:inverse}
\end{eqnarray}
Consequently, in the approximation of reflectless potentials ($R(k) = 0$),
the wave functions, the single-particle potential and the density profiles 
are completely
defined by the bound state eigenvalues via formulas 
(\ref{eq:slab}),(\ref{eq:inverse}).
%
%
\section{Results and Discussion} 
In Ref. \cite{kartmad87} a series of calculations for the 
different layers, imitating nuclear
systems in their ground state was provided.
For a direct part of the calculations by the Hartree-Fock method, the
interaction functional was chosen in the form of effective Skyrme forces. 
The calculated spectrum of bound states was fed into the scheme of the inverse
scattering method, and the relations 
were used to recover the wave functions of
the states, the single-particle  potentials, and the corresponding densities.
In this note, we generalize this method to the case of 
fragmented  configuration, trying to imitate two- and three-center 
nuclear systems. We use here only the inverse mean-field scheme
(\ref{eq:inverse}). The details of the approach and
systematic calculations of fragmented nuclear systems will be
provided in a forthcoming publication.
%
\begin{figure}[h]
\begin{center}
\includegraphics[width=150mm]{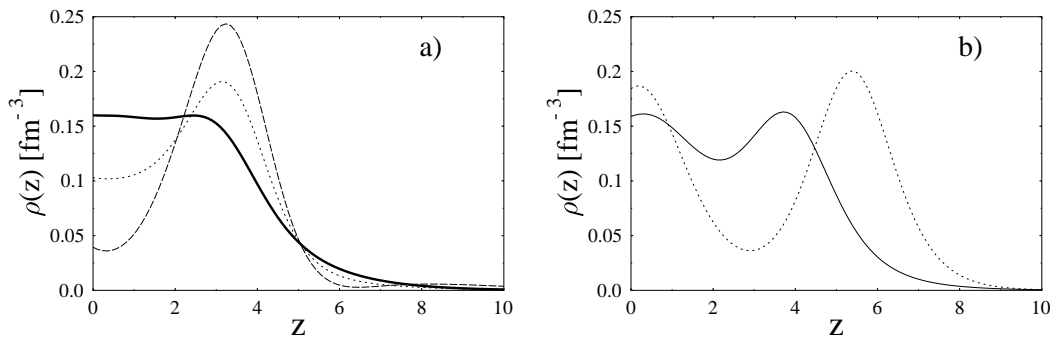}   
\end{center}
\vspace*{-5mm}
\caption{The density profiles of the light 
($A\approx 20,\; {\cal A}\approx~1.0$)
three-levels ($N=3,\; N_0=3$)      
model system calculated in the frame of inverse mean field method.
a) the ground state (solid line); a fragmented two-center configuration
(dotted and dashed lines);
b) a ternary fragmentation of the system into three fragments.  
}
\label{fig}
\end{figure}
%
In Fig.1 we present the results of the calculations of three-level
($N=3,\; N_0=3$) light ($A\approx 20,\; {\cal A}\approx~1.0$)     
model system simulating the ground state
(Fig.~\ref{fig}(a) solid line), and
fragmentation of the system into two fragments (Fig.~\ref{fig}(a), 
the dotted and dashed lines).
In the same figure (Fig.~\ref{fig}(b)) we  present the 
fragmentation of the system into three fragments (solid and dotted 
lines).
One can see that it is possible to simulate 
a one-dimensional three-center system 
via inverse scattering method.
The following conclusions can be drawn.
\begin{itemize}
\item{}
The density profiles,
calculated in the framework of inverse method, are practically
identical to the results of calculation by SHF
method. These results are valid for the ground state and 
for the system in the external potential field.  
\item{}
The global properties of single-particle potentials 
(the depth and an effective radius)
have been reproduced quite well, but the 
inverse method yields 
the quite strongly pronounced oscillations of the potential
distributions within the inner region, and
slightly different asymptotic tails of potential.
In the framework of inverse scattering method, all bound states are taken into
account in the calculation of the potential (\ref{eq:inverse}), but for the
density distribution only the occupied states are taken into account 
(see Eqs.~(\ref{eq:density})). 
Therefore, the slope of the tails of the potential and of
the density distributions will we different. 
\item{}
We used, the approximation of reflectionless potentials, which gave us the
possibility to obtain a simple closed set of relations (\ref{eq:inverse}),
to calculate wave functions, density distributions and single particle
potentials. 
The omitted reflection terms ($R(k)=0$)are not important 
for the evaluation of the density
distributions, due to the fact that only
the deepest occupied states are used to evaluate density distribution
(see Eq.~(\ref{eq:density})).
The introduction of these reflection terms will lead to a smoothing of the
oscillations in the inner part of the potential and to a correction of its
asymptotic behaviour.  
\item{}
The presented method gives a tool to simulate the  various sets of the
static excited states of the system.
This method could be useful to
prepare in a simple way an initial state for the 
dynamical calculations in the frame of mean-field methods. 
\end{itemize}
\section{Conclusions}
Recent experimental progress in the investigation of cold nuclear 
fragmentation has made the development of theoretical many-body methods 
highly desirable. 
Modern variants of self-consistent Hartree-Fock and 
relativistic mean-field models give the principal way 
to describe nuclear fragmentation in the
framework of many-body self-consistent approach.
However, the generalization of these approaches to three-center
case is not provided yet because of existing difficulties to select a
suitable set of constraints.

We applied the inverse scattering method
to solve the single-particle Schr\"odinger equation,
instead of direct solution of constrained
self--consistent Hartree--Fock equations.
It is shown,
that it  is possible to simulate the one-dimensional three-center system 
via inverse scattering method.

It is needless to say that the present one-dimensional 
model is too primitive
to describe a real three-dimensional fragmentation in nuclear systems.
However this model may be useful as a guide to understand
the general properties
of fragmented systems and to formulate the suitable set of constraints
for the realistic three-dimensional mean field calculations of the
three-center nuclear system. 
%
%
%

\eject
\section*{References}

\end{document}